\documentclass[pra,onecolumn,showpacs,superscriptaddress,amsmath,amssymb]{revtex4} 
\usepackage{graphicx,bm} 
\usepackage{color,ulem} 

\begin{document} 

\title{Hydrodynamic equation of a spinor dipolar Bose-Einstein condensate} 
\author{Kazue Kudo} 
\affiliation{Division of Advanced Sciences, Ochadai Academic Production, 
Ochanomizu University, 2-1-1 Ohtsuka, Bunkyo-ku, Tokyo 112-8610, Japan} 
\author{Yuki Kawaguchi} 
\affiliation{Department of Physics, University of Tokyo, 7-3-1 Hongo,
Bunkyo-ku, Tokyo 113-0033, Japan} 
 
\date{\today} 
\begin{abstract} 
We introduce equations of motion for spin dynamics in a
ferromagnetic Bose-Einstein condensate with magnetic dipole-dipole
 interaction, written using a vector
expressing the superfluid velocity and a complex scalar describing the
 magnetization. This simple hydrodynamical description extracts the 
dynamics of spin wave and affords a
 straightforward approach by which to
 investigate the spin dynamics of the condensate. 
To demonstrate the advantages of the description,
we illustrate dynamical instability and magnetic
 fluctuation preference, which are expressed in analytical forms.
\end{abstract} 
\pacs{03.75.Lm, 03.75.Mn,03.75.Kk} 
\maketitle 
 
\section{Introduction}

One of the salient features of a gaseous Bose-Einstein condensate (BEC) 
is the internal spin degrees of freedom.
In spinor BECs, namely, in BECs with internal degrees of freedom, 
spin and gauge degrees of freedom couple in various manners,
leading to nontrivial properties of spin waves and topological excitations.
For example, ferromagnetic BECs have continuous spin-gauge symmetry,
thus the circulation of the superfluid velocity is not
quantized~\cite{Ho1996,Nakahara2000,Leanhardt2003}, 
whereas spin-1 polar BECs and spin-2 cyclic BECs can host fractional 
vortices due to the discrete spin-gauge 
symmetry~\cite{Makela,Zhou2001,Semenoff2007,Kobayashi2009}. 
In recent experiments,
{\it in situ} imaging of transverse magnetization has revealed the
real-time dynamics of the spontaneous symmetry breaking, spin texture
formation, 
and nucleation of spin 
vortices~\cite{Sadler2006, berkeley08, Vengalattore2010}, 
opening up a new paradigm for studying the static and dynamic properties 
of spin textures. 

On the other hand, BECs with magnetic dipole-dipole interaction (MDDI)
have also attracted much attention both experimentally and theoretically in
recent years. 
The long-range and anisotropic nature of the MDDI is predicted to yield
exotic phenomena, such as new equilibrium shapes, roton-maxon spectra,
supersolid states, and two-dimensional solitons~\cite{MDDI_review}. 
In particular, when the BEC has spin degrees of freedom,
the MDDI is predicted to develop spin 
textures, even when the MDDI is much weaker than the contact
interaction~\cite{Santos2006,Kawaguchi2006,Yi2006}. 
This work is motivated by experiments done by the Berkeley 
group~\cite{berkeley08}, where small magnetic domains were observed 
to develop from a helical
spin structure in a spin-1 $^{87}$Rb BEC. The method presented 
in this paper
simplifies the spin dynamics in a complicated system of spinor
dipolar BECs, although we have shown in our previous work 
that mean-field calculations do
not reproduce the experimental results~\cite{kawa09}.

In this paper, we propose a new type of hydrodynamic description of a
ferromagnetic BEC with MDDI. 
The hydrodynamic equation of spinor BECs has been discussed for both
ferromagnetic phases~\cite{Takahashi2007, lama, Barnett2009} 
and non-magnetized phases~\cite{Barnett2009,Lamacraft2010}.
In Ref.~\cite{Takahashi2007}, Takahashi \textit{et al.} consider 
the strong MDDI limit by using the
classical spin model, i.e., by neglecting the spin-gauge coupling. 
On the other hand, Lamacraft
takes into account the spin-gauge
coupling by introducing the so-called Mermin-Ho relation, 
and considers the weak MDDI~\cite{lama}.
In these papers, the authors use a unit vector to describe the
local magnetization in the ferromagnetic phase. 
Here we use a single complex scalar variable instead of a unit vector to
describe the local magnetization and treat both the 
spin-gauge coupling and MDDI.  
This simple description allows a straightforward approach to
analyze the spin dynamics of the condensate.
In order to demonstrate the advantages of our description, 
we analyze the dynamical instability and magnetization fluctuation
preference of the BEC with MDDI.

The rest of the paper is organized as follows.
In Sec.~\ref{sec.hydro}, hydrodynamic equations described using the spin
density vector are derived from the Gross-Pitaevskii (GP)
equation. We then rewrite the equations by means of 
stereographic projection for some simple cases: 
quasi two-dimensional (2D) systems under zero external field and
under a strong magnetic field. 
For both zero-field and strong-field cases, 
the wavevector dependence of the dynamical instability is
obtained straightforwardly in an analytical form 
in Sec.~\ref{sec.instability}. 
In Sec~\ref{sec.fluctuation}, we also illustrate the
magnetic fluctuation preference for the unstable modes discussed in
Sec.~\ref{sec.instability}.  Conclusions are given in
Sec.~\ref{sec.conc}. 

\section{\label{sec.hydro} Hydrodynamic description}

\subsection{Equations of motion of mass and spins}

We consider a spin-$F$  BEC  under a
uniform magnetic field 
$B$ applied in the $z$ direction. 
The  GP equation for the spinor dipolar system is given by
\begin{eqnarray}
i\hbar \frac{\partial}{\partial t} \Psi_m(\bm{r},t) 
&=& 
(H_0 + pm + qm^2) \Psi_m \nonumber\\
&& + \sum_{S=0,{\rm even}}^{2F} \frac{4\pi\hbar^2}{M}a_S  \sum_{M_S=-S}^S 
 \sum_{n, m',n'=-F}^{F} 
\langle mn|S M_S\rangle \langle S M_S | m'n'\rangle
\Psi_n^* \Psi_{m'} \Psi_{n'}
\nonumber \\
&&
+ c_{\rm dd} \sum_{\mu =x,y,z} \sum_{n=-F}^{F} b_\mu
(F_\mu)_{mn} \Psi_n,
\label{eq.GP}
\end{eqnarray}
where $\Psi_m(\bm{r},t)$ is the condensate wavefunction for the atoms in the
magnetic sublevel $m$ and
$H_0 = -\hbar^2\nabla^2/(2M) + U_{\rm trap}(\bm{r})$,
with $M$ being the atomic mass and $U_{\rm trap}$ the spin-independent
trapping potential. 
The linear and quadratic Zeeman energies per atom are
given by  $p = g_F \mu_{\rm B}B$ and
$q = (g_F \mu_{\rm B} B)^2/E_{\rm hf}$, respectively, 
where $g_F$ is the hyperfine $g$-factor, 
$\mu_{\rm B}$ is the Bohr magneton, and
$E_{\rm hf}$ is the hyperfine energy splitting.
The second term on the right-hand side of Eq.~\eqref{eq.GP} comes from
the short-range part of the two-body interaction given by 
\begin{align}
V_{\rm s}(\bm r,\bm r') = \delta(\bm r-\bm r')
\sum_{S=0,{\rm even}}^{2F}\frac{4\pi\hbar^2}{M}a_S\mathcal{P}_S, 
\label{eq.Vs}
\end{align}
where $\mathcal{P}_S=\sum_{M_S=-S}^S|S M_S\rangle \langle S M_S|$
projects a pair of spin-1 atoms into the state with total spin $S$, 
and $a_S$ is the $s$-wave scattering length for the corresponding spin
channel $S$. 
The scattering amplitude for odd $S$ vanishes due to Bose symmetrization,
and $\langle mn|S M_S\rangle$ in Eq.~\eqref{eq.GP} is the Clebsch-Gordan
coefficient. 
The last term on the right-hand side of Eq.~\eqref{eq.GP} corresponds to
the MDDI, where 
$c_{\rm dd} = \mu_0(g_F \mu_{\rm B})^2/(4\pi)$,
with $\mu_0$ being the magnetic permeability of the vacuum.
Here, we define the non-local dipole field by
\begin{equation}
 b_\mu(\bm{r}) =\int d^3 r' \sum_\nu Q_{\mu\nu}
(\bm{r}-\bm{r}') f_\nu(\bm{r}'),
\end{equation}
where 
$Q_{\mu\nu}(\bm{r})$
is the dipole kernel, given in Sec.~\ref{sec:2d}, and
\begin{align}
f_\mu = \sum_{mn} \Psi^*_m(F_\mu)_{mn}\Psi_n
\end{align}
is the spin density, with $F_{x,y,z}$ being the spin-$F$ matrices.
Below we omit the summation symbol: Greek indices that appear twice
are to be summed over $x$, $y$, and $z$, 
and Roman indices are to be summed over $-F,\ldots,F$. 

From the GP equation, we can immediately derive the mass continuity equation:
\begin{equation}
 \frac{\partial n_{\rm tot}}{\partial t} + \nabla\cdot  
\left(n_{\rm tot}\bm{v}_{\rm mass}\right) = 0,
\label{eq.cont_mass}
\end{equation}
where 
\begin{align}
 n_{\rm tot} &= \Psi_m^* \Psi_m, \\
 n_{\rm tot}\bm{v}_{\rm mass} &= \frac{\hbar}{2Mi} 
[ \Psi_m^*(\nabla\Psi_m) - (\nabla\Psi_m^*)\Psi_m]
\end{align}
are the number density and superfluid current, respectively.
By introducing a normalized spinor $\zeta_m$ defined by 
$\Psi_m(\bm r, t)=\sqrt{n_{\rm tot}(\bm r, t)}\zeta_m(\bm r,t)$,
the superfluid velocity $\bm{v}_{\rm mass}$ can be written as
\begin{align}
 \bm v_{\rm mass}&= \frac{\hbar}{2Mi} 
[ \zeta_m^*(\nabla\zeta_m) - (\nabla\zeta_m^*)\zeta_m].
\label{eq.v_mass_def}
\end{align}
In the absence of the linear and quadratic Zeeman effects and the MDDI, 
the continuity equation of the spin density can also be derived from
the GP equation as
\begin{equation}
 \frac{\partial f_\mu}{\partial t} + \nabla\cdot 
\left(n_{\rm tot}\bm{v}^\mu_{\rm spin}\right) =0,
\label{eq.cont_spin}
\end{equation}
where 
$\bm{v}_{\rm spin}^\mu$ is the spin superfluid velocity defined by
\begin{equation}
  \bm{v}^\mu_{\rm spin} = \frac{\hbar}{2Mi} (F_\mu)_{mn}
[\zeta^*_m(\nabla\zeta_n) - (\nabla\zeta_m^*)\zeta_n].
\label{eq.v_spin_def}
\end{equation}
The short-range interaction does not contribute to the equation of 
motion of spin, 
since it conserves the total spin of two colliding atoms.
The detailed calculation is given in Appendix~\ref{sec.A}.
In the presence of the external magnetic field along the $z$ direction,
the linear Zeeman effect induces a torque term  
$(p/\hbar)(\hat{z}\times\bm{f})_\mu$
on the right-hand side of Eq.~(\ref{eq.cont_spin}),
which  causes the precession of spins.
In a similar manner, 
the dipole field also induces a torque term
$(c_{\rm dd}/\hbar)(\bm{b}\times\bm{f})_\mu$.
On the other hand, the quadratic Zeeman term does not conserve the
transverse magnetization and its effect is written as 
$(2q/\hbar)\epsilon_{\mu z \nu} n_{\rm tot}\hat{\mathcal{N}}_{z\nu}$, where 
$\epsilon_{ijk}$ is the Levi-Civita symbol and
\begin{align}
\hat{\mathcal{N}}_{\mu\nu} 
= \frac12 \zeta_m^*(F_\mu F_\nu + F_\nu F_\mu)_{mn}\zeta_n
\label{eq.nematic}
\end{align}
is a nematic tensor.
The derivations of these three terms are given in Appendix~\ref{sec.A}. 
As a result, we obtain the equation of motion of spins in the presence
of the MDDI and linear and quadratic Zeeman effects  
under the external field parallel to the $z$ axis:
\begin{equation}
 \frac{\partial f_\mu}{\partial t} 
+ \nabla\cdot\left( n_{\rm tot}\bm{v}^\mu_{\rm spin}\right)
= \frac{c_{\rm dd}}{\hbar} (\bm{b} \times \bm{f})_\mu
+ \frac{p}{\hbar}(\hat{z}\times\bm{f})_\mu
+ \frac{2q}{\hbar}n_{\rm tot} \epsilon_{\mu z \nu}
\hat{\mathcal{N}}_{z\nu}.
\label{f.0}
\end{equation}
Equations~\eqref{eq.cont_mass} and \eqref{f.0} hold in all phases,
independent of scattering length. 

\subsection{Ferromagnetic BEC}

In the following,
we consider a ferromagnetic BEC.
We assume that the BEC is fully magnetized, $|\bm f|=Fn_{\rm tot}$, and
only the direction of the spin density can vary in space. 
This assumption is valid 
when the ferromagnetic interaction energy is large enough compared with 
the other
spinor interaction energies, MDDI energy, quadratic Zeeman energy, 
and the kinetic energy arising from 
the spacial variation of the direction of $\bm f$.
The linear Zeeman effect is not necessarily weaker than
the ferromagnetic interaction, since it merely induces the Larmor precession.
For example, the short-range interaction~\eqref{eq.Vs} for a spin-1 BEC
can be written as~\cite{Ho1998} 
\begin{align}
 \langle mn |V_{\rm s}(\bm r, \bm r') |m'n'\rangle 
= \delta(\bm r-\bm r') \left[c_0 \delta_{mn}\delta_{m'n'} 
+ c_1 (F_\mu)_{mn}(F_\mu)_{m'n'}\right],
\end{align}
where $c_0=4\pi\hbar^2(2a_2+a_0)/(3M)$ and $c_1=4\pi\hbar^2(a_2-a_0)/(3M)$.
The ground state is ferromagnetic for $c_1<0$.
The above assumption is valid when $q\ll |c_1|n_{\rm tot}$,
$c_{\rm dd}\ll |c_1|$, and
the length scale of the spatial spin structure is larger than the spin
healing length $\xi_{\rm sp}=\hbar/\sqrt{2M|c_1|n_{\rm tot}}$. 
Moreover in the incompressible limit, namely when the spin independent
interaction ($c_0 n_{\rm tot}$ for the case of a spin-1 BEC) is much
stronger than the ferromagnetic interaction and MDDI, 
the number density $n_{\rm tot}$ is determined regardless of the spin
structure and assumed to be stationary.
This is the case for the spin-1 $^{87}$Rb BEC.

\begin{figure}[tb]
\includegraphics[width=4cm]{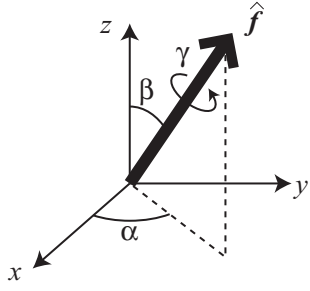}
\caption{\label{fig.E} Euler rotation of the unit vector $\hat{{\bm f}}$.}
\end{figure}

We then rewrite the equations of motion~\eqref{eq.cont_mass} and \eqref{f.0}
in terms of a unit vector $\hat{\bm f}\equiv {\bm f}/(Fn_{\rm tot})$
that describes the direction of the spin density 
and the superfluid velocity $\bm{v}_{\rm mass}$ defined in
Eq.~\eqref{eq.v_mass_def}. 
The order parameter for the spin-polarized state in the $z$ direction is
given by $\zeta^{(0)}_m=\delta_{mF}$. 
The general order parameter is obtained by performing the gauge
transformation and Euler rotation as 
\begin{align}
 \bm \zeta &= e^{i\phi}e^{-iF_z\alpha}e^{-iF_y\beta}e^{-iF_z\gamma}
\bm \zeta^{(0)} \nonumber\\
 &= e^{i(\phi-F\gamma)}e^{-iF_z\alpha}e^{-iF_y\beta}\bm \zeta^{(0)},
\label{eq.zeta_general}
\end{align}
where $\alpha$, $\beta$ and $\gamma$ are Euler angles shown in
Fig.~\ref{fig.E} and $\phi$ is the overall phase. 
Due to the spin-gauge symmetry of the ferromagnetic BEC, i.e.,
the equivalence between the phase change $\phi$ and spin rotation $\gamma$,
distinct configurations of $\bm \zeta$ are characterized with a set
of parameters $(\alpha, \beta, \phi'\equiv\phi-F\gamma)$. 
Here, $\alpha$ and $\beta$ denote the direction of $\hat{\bm f}$ as
shown in Fig.~\ref{fig.E}. 
Actually, $\hat{\bm f}$ for the order parameter~\eqref{eq.zeta_general}
is calculated as 
\begin{align}
 F\hat{\bm f} &= \zeta_m^* {\bm F}_{mn} \zeta_n \nonumber\\
 &= \zeta_m^{(0)*} \left(e^{iF_y\beta} e^{iF_z\alpha} {\bm F} 
e^{-iF_z\alpha} e^{-iF_y\beta}\right)_{mn} \zeta_n^{(0)} \nonumber\\
 &= \mathcal{R}_z(\alpha) \mathcal{R}_y(\beta) 
\left[\zeta_m^{(0)*}{\bm F}_{mn} \zeta_n^{(0)}\right] \nonumber\\
 &= F\begin{pmatrix} 
\sin\beta\cos\alpha \\ \sin\beta\sin\alpha \\ \cos\beta 
\end{pmatrix},
\end{align}
where $\mathcal{R}_z(\alpha)$ and $\mathcal{R}_y(\beta)$ are the
$3\times 3$ matrices describing the rotation 
about the $z$ axis by $\alpha$ and about the $y$ axis by $\beta$,
respectively. 
In a similar manner, we obtain the nematic tensor
$\hat{\bm{\mathcal{N}}}$ for the order parameter~\eqref{eq.zeta_general}
as 
\begin{align}
 \hat{\mathcal{N}}_{\mu\nu} &= \mathcal{R}_z(\alpha)
 \mathcal{R}_y(\beta) \hat{\mathcal{N}}_{\mu\nu}^{(0)} 
\mathcal{R}^{\rm T}_y(\beta)\mathcal{R}^{\rm T}_z(\alpha) \nonumber\\
 &=\frac{F}{2}\delta_{\mu\nu} + \frac{F(2F-1)}{2}\hat{f}_\mu \hat{f}_\nu,
\label{eq.nematic_ferro}
\end{align}
where T denotes the transpose and 
$\hat{\mathcal{N}}_{\mu\nu}^{(0)}=\frac12 \zeta_m^{(0)*}(F_{\mu}F_{\nu}
+ F_{\nu}F_{\mu})_{mn}\zeta_n^{(0)}$ is
the nematic tensor for $\bm\zeta^{(0)}$, which is given by 
\begin{align}
 \hat{\bm{\mathcal{N}}}^{(0)} &=\begin{pmatrix} 
F/2 & 0 & 0 \\ 0 & F/2 & 0 \\ 0 & 0 & F^2 
\end{pmatrix} \nonumber\\
 &= \frac{F}{2} \begin{pmatrix} 
1 & 0 & 0 \\ 0 & 1 & 0 \\ 0 & 0 & 1 
\end{pmatrix} 
 + \frac{F(2F-1)}{2} \begin{pmatrix} 
0 & 0 & 0 \\ 0 & 0 & 0 \\ 0 & 0 & 1 
\end{pmatrix}.
\end{align}
Substituting Eq.~\eqref{eq.zeta_general} and 
\begin{align}
 \nabla \bm\zeta = i\left[\nabla \phi' - (\nabla\alpha)F_z 
- (\nabla \beta) e^{-iF_z \alpha}F_y e^{iF_z\alpha}\right]\bm \zeta
\label{eq.nabla_zeta}
\end{align}
into Eq.~\eqref{eq.v_mass_def}, the superfluid velocity can be written as
\begin{equation}
 \bm{v}_{\rm mass} = \frac{\hbar}{M} 
[\nabla\phi' - F(\nabla\alpha)\cos\beta],
\label{v_mass.0}
\end{equation}
which satisfies the Mermin-Ho relation \cite{Mermin}:
\begin{equation}
 \nabla \times \bm{v}_{\rm mass} = \frac{\hbar F}{2M}
\epsilon_{\mu\nu\lambda} \hat{f}_\mu 
( \nabla\hat{f}_\nu \times \nabla\hat{f}_\lambda ).
\label{eq.MHR}
\end{equation}
As we mentioned before, 
$n_{\rm tot}$ is stationary in the incompressible limit. Thus,
Eq.~\eqref{eq.cont_mass} becomes 
\begin{equation}
\nabla\cdot (n_{\rm tot}\bm{v}_{\rm mass}) = 0.
\label{div.0}
\end{equation}
Equations~\eqref{eq.MHR} and \eqref{div.0} are equations for the
superfluid velocity. 
Next, we consider the equation for the spin superfluid velocity.
Making use of Eq.~\eqref{eq.nematic_ferro},
Eq.~\eqref{eq.v_spin_def} can be rewritten in terms of $\hat{\bm f}$ and
$\bm v_{\rm mass}$ as 
\begin{align}
 \bm{v}_{\rm spin}^\mu &= \frac{\hbar}{M} 
\left[ (\nabla \phi')F\hat{f}_\mu  - (\nabla \alpha)\hat{\mathcal{N}}_{\mu z} 
 - (\nabla \beta)\hat{\mathcal{N}}_{\mu y} \cos\alpha + (\nabla\beta)
 \hat{\mathcal{N}}_{\mu x} \sin\alpha \right] \nonumber\\
 &= F\hat{f}_\mu \bm{v}_{\rm mass} - \frac{\hbar F}{2M} \left[ 
 (\nabla\alpha)\sin\beta \begin{pmatrix} 
-\cos\beta \cos\alpha \\ -\cos\beta\sin\alpha \\ \sin\beta 
\end{pmatrix}
 + (\nabla\beta)\begin{pmatrix} 
-\sin\alpha \\ \cos\alpha \\ 0 
\end{pmatrix}
\right]_\mu \nonumber\\
 &=  F\hat{f}_\mu \bm{v}_{\rm mass} - \frac{\hbar F}{2M} 
\epsilon_{\mu\nu\lambda} \hat{f}_\nu \nabla \hat{f}_\lambda.
\label{v.spin}
\end{align}
Substituting Eqs.~\eqref{eq.nematic_ferro}, \eqref{div.0}, and
\eqref{v.spin} into Eq.~\eqref{f.0}, 
we obtain the hydrodynamic equation in terms of $\hat{\bm{f}}$ and
$\bm{v}_{\rm mass}$ as
\begin{equation}
    \frac{\partial \hat{\bm{f}}}{\partial t} 
+ (\bm{v}_{\rm mass}\cdot\nabla)\hat{\bm{f}}
= -\hat{\bm{f}}\times\bm{B}_{\rm eff},
\label{f.1}
\end{equation}
with
\[
 \bm{B}_{\rm eff} = -\frac{\hbar}{2M}(\bm{a}\cdot\nabla)\hat{\bm{f}}
-\frac{\hbar}{2M}\nabla^2\hat{\bm{f}}
+ \frac{c_{\rm dd}}{\hbar}\bm{b} + \frac{p}{\hbar}\hat{e}^B
+ \frac{q(2F-1)}{\hbar}(\hat{e}^B\cdot\hat{\bm{f}})\hat{e}^B,
\]
where
$\bm{a}=(\nabla n_{\rm tot})/n_{\rm tot}$ and 
$\hat{e}^B$ is the unit vector along the external field 
($\hat{e}^B = \hat{z}$ in this paper). 
Here we note that Eq.~(\ref{f.1}) has the same form as the extended 
Landau-Lifshitz-Gilbert equation (without damping) 
which includes the adiabatic spin torque term~\cite{LLG}.

\subsection{\label{sec:2d}Quasi-2D system}

We next consider a quasi-2D system, that is, we consider a BEC
confined in a quasi-2D trap whose Thomas-Fermi radius in the normal
direction to the 2D plane is smaller than the spin healing length. 
We approximate the wavefunction in the normal direction
by a Gaussian with width $d$: 
$\Psi_m(\bm{r}_\perp,r_{\rm n}) = \psi_m(\bm{r}_\perp)h(r_{\rm n})$, where
$\bm{r}_\perp$ is the position vector in the 2D plane, 
$r_{\rm n}$ is the coordinate in the normal direction,
and $h(r_{\rm n}) = \exp[-r_{\rm n}^2/(4d^2)]/(2\pi d^2)^{1/4}$. 
Multiplying the wavefunction to Eq.~(\ref{eq.GP}) and
integrating over $r_{\rm n}$, 
we obtain the 2D GP equation.
The equation is the same as Eq.~\eqref{eq.GP} if one replaces 
$\Psi_m$ with $\psi_m$, 
$a_S$ with $\eta a_S$, $c_{\rm dd}$ with $\eta c_{\rm dd}$, and $\bm{b}$
with $\bar{\bm{b}}$, 
where $\eta = \int dr_{\rm n} h^4(r_{\rm n})/\int dr_{\rm n} h^2(r_{\rm n}) = 1/\sqrt{4\pi d^2}$
and
\begin{align}
 \bar{b}_\mu = \int d^2r'_\perp Q_{\mu\nu}^{\rm (2D)}
(\bm r_\perp - \bm r'_\perp) 
\left[\psi_m^*(\bm r_\perp')(F_\nu)_{mn}\psi_n(\bm r_\perp')\right],
\label{eq:b_bar}
\end{align}
with
\begin{align}
 Q^{\rm (2D)}_{\mu\nu}({\bm r}_\perp-{\bm r}_\perp') 
= \frac{1}{\eta} \iint dr_{\rm n} dr'_{\rm n} 
h^2(r_{\rm n})h^2(r_{\rm n}') Q_{\mu\nu}({\bm r}-{\bm r}').
\label{eq:Q_3D_to_2D}
\end{align}
Starting from the 2D GP equation and following the above procedure, we
derive the 2D hydrodynamic equation: 
\begin{equation}
    \frac{\partial \hat{\bm{f}}}{\partial t} 
+ (\bm{v}_{\rm mass}\cdot\nabla)\hat{\bm{f}}
= -\hat{\bm{f}}\times\bar{\bm{B}}_{\rm eff},
\label{f.1.2D}
\end{equation}
with
\[
 \bar{\bm{B}}_{\rm eff} = -\frac{\hbar}{2M}(\bm{a}\cdot\nabla)\hat{\bm{f}}
-\frac{\hbar}{2M}\nabla^2\hat{\bm{f}}
+ \frac{\eta c_{\rm dd}}{\hbar}\bar{\bm{b}} + \frac{p}{\hbar}\hat{e}^B
+ \frac{q(2F-1)}{\hbar}(\hat{e}^B\cdot\hat{\bm{f}})\hat{e}^B,
\]
where $\bm{v}_{\rm mass}$ and $\nabla$ are the two-dimensional 
vector and vector operator, respectively.
When we consider a quasi-2D BEC,
$n_{\rm tot}$, $\hat{\bm{f}}$, and $\bm{v}_{\rm mass}$ are defined
by means of $\psi_m$ 
instead of $\Psi_m$. 

\subsection{Dipole kernel}

This section provides the detailed form of the dipole kernel in 3D and
quasi-2D systems under zero external field and under 
a strong magnetic field 
($p\gg c_{\rm dd}n_{\rm tot}$). 
The derivations are given in Ref.~\cite{kawa09}. 

The dipole kernel in the laboratory frame of reference is given by
\begin{align}
Q^{\rm (lab)}_{\mu\nu}({\bm r}) 
&= \frac{\delta_{\mu\nu}-3\hat{r}_\mu\hat{r}_\nu}{r^3},
\label{eq:Q_lab}
\end{align}
with $r=|\bm r|$ and $\hat{\bm r}=\bm r/r$.
The 2D dipole kernel in the laboratory frame 
is calculated by substituting Eq.~\eqref{eq:Q_lab}
into Eq.~\eqref{eq:Q_3D_to_2D}. 
For $\bm r_\perp=(x,y)$ and $r_{\rm n}=z$, the 2D dipole kernel is given by
\begin{align}
 Q^{\rm (2D,lab)}_{\mu\nu} ({\bm r}_\perp) &= \sum_{\bm k_\perp} 
e^{i\bm k_\perp\cdot \bm r_\perp}
\tilde{Q}^{\rm (2D,lab)}_{\bm k_\perp\mu\nu},
\label{Q2Dlab}
\end{align}
where
\begin{align}
 \tilde{Q}^{\rm (2D,lab)}_{\bm{k}_\perp}
&= 
- \frac{4\pi}{3} \left(
\begin{array}{ccc}
 1 & 0 & 0 \\
 0 & 1 & 0 \\
 0 & 0 & -2
\end{array}
\right)
+ 4\pi G(k_\perp d) \left(
\begin{array}{ccc}
 \hat{k}_x^2 & \hat{k}_x\hat{k}_y & 0 \\
 \hat{k}_x\hat{k}_y & \hat{k}_y^2 & 0 \\
 0 & 0 & -1
\end{array}
\right),
\label{Qk.A}
\end{align}
with $\bm{k}_\perp = (k_x,k_y)$, $k_\perp = |\bm{k}_\perp|$,
$\hat{k}_{x,y} = k_{x,y}/k_\perp$,
and 
$G(k) \equiv 2 k e^{k^2} \int_k^\infty e^{-t^2}dt = \sqrt{\pi}ke^{k^2}\mathrm{erfc}(k)$.
It can be shown that $G(k)$ is a monotonically increasing function
that satisfies $G(0)=0$ and $G(\infty)=1$.

When the linear Zeeman energy is much larger than the MDDI energy,
we choose the rotating frame of reference in spin space by replacing
$\Psi_m$ with $e^{-ipmt/\hbar}\Psi_m$, 
and eliminate the linear Zeeman term from the GP equation.
In this case, the contribution of the MDDI is time-averaged due to the
Larmor precession, 
and we use the dipole kernel which is averaged over the Larmor precession
period given by~\cite{kawa07} 
\begin{align}
Q_{\mu\nu}^{\rm (rot)}({\bm r}) 
&= -\frac{1}{2}\,\frac{1-3\hat{r}_z^2}{r^3}
\left(\delta_{\mu\nu}-3\delta_{z\mu}\delta_{z\nu}\right).
\label{eq:Q_rot}
\end{align}
Substituting Eq.~\eqref{eq:Q_rot} into Eq.~\eqref{eq:Q_3D_to_2D}, we
obtain the time-averaged 2D dipole kernel in the rotating frame as 
\begin{align}
 Q^{\rm (2D, rot)}_{\mu\nu}(\bm r_\perp)
  &=  \left(\delta_{\mu\nu}-3\delta_{z\mu}\delta_{z\nu}\right)
\sum_{\bm k_\perp}e^{i\bm k_\perp\cdot \bm r_\perp}
\tilde{\mathcal{Q}}_{{\bm k}_\perp},
\label{eq:Q_rot_2D_k}
\end{align}
where
\begin{align}
  \tilde{\mathcal{Q}}_{{\bm k}_\perp}
=\frac{2\pi}{3}\left\{1-3(\hat{e}_{\rm n}\cdot \hat{e}^B)^2 
 -3G(k_\perp d)\left[(\hat{\bm e}_\perp^B\cdot\hat{\bm k}_\perp)^2 
- (\hat{e}_{\rm n}\cdot\hat{e}^B)^2 \right]\right\}.
\label{Qk.B}
\end{align}
Here, $\hat{\bm{k}}_\perp = \bm{k}_\perp / k_\perp$,
$\hat{e}_{\rm n}$ is the unit vector normal to the plane, and
$\hat{\bm{e}}^B_\perp$ is the vector of $\hat{e}^B$ projected onto the
2D plane.

\subsection{Stereographic projection}

The spin dynamics are now described by Eqs.~(\ref{eq.MHR}), (\ref{div.0}),
 and (\ref{f.1}) or (\ref{f.1.2D}).
We rewrite the equations by means of
stereographic projection \cite{laksh}: we employ a complex number
$\varphi=(\hat{f}_x + i\hat{f}_y)/(1+\hat{f}_z)$ to express the spin
variables, 
\begin{equation}
 \hat{f}_x = \frac{\varphi + \varphi^*}{1 + \varphi\varphi^*}, \quad
 \hat{f}_y = \frac{-i(\varphi - \varphi^*)}{1 + \varphi\varphi^*}, \quad
 \hat{f}_z = \frac{1 - \varphi\varphi^*}{1 + \varphi\varphi^*}.
\label{f_xyz}
\end{equation}
Equation~(\ref{eq.MHR}) is rewritten as 
\begin{equation}
 \nabla\times\bm{v}_{\rm mass} = i F \frac{2\hbar}{M}
\frac{\nabla\varphi \times \nabla\varphi^*}{(1+\varphi\varphi^*)^2},
\label{eq.MHR.1}
\end{equation}
while Eq.~(\ref{div.0}) remains the same. 
In order to rewrite the equation of spins, we need to specify the
dimensionality of the system and the direction and strength of the
external field. In this paper, we consider the following two cases: 
(i) a quasi-2D system normal to the $z$ axis under zero magnetic field,
and (ii) a quasi-2D system normal to the $y$ axis with a strong magnetic
field along the $z$ axis. Case (ii) corresponds to the situation in
the Berkeley experiment~\cite{berkeley08}. 

For case (i), we take $\hat{e}_{\rm n}=\hat{z}$ and $p=q=0$ and use
Eqs.~(\ref{eq:b_bar}), (\ref{f.1.2D}), (\ref{Q2Dlab}), and (\ref{Qk.A}). 
Using the
stereographic projection, the equation of motion of spins is given by 
\begin{eqnarray}
 \frac{\partial\varphi(\bm{r},t)}{\partial t} 
&=& 
-\bm{v}_{\rm mass}\cdot\nabla\varphi
+ \frac{i\hbar}{2Mn_{\rm tot}}\nabla n_{\rm tot}\cdot\nabla\varphi
+ \frac{i\hbar}{2M}\nabla^2\varphi
- \frac{i\hbar}{M} \frac{\varphi^*(\nabla\varphi)^2}{1+\varphi\varphi^*}
\nonumber \\
&&
- \frac{i\eta c_{\rm dd}F}{2\hbar} \int d^2 r' 
n_{\rm tot}(\bm{r}') \sum_{\bm{k}} e^{i\bm{k}\cdot (\bm{r}-\bm{r}')}
\left[ 
- h_1(k)
\frac{\varphi(\bm{r}')}{1 + \varphi(\bm{r}')\varphi^*(\bm{r}')}
+ h_2(k)
\frac{\varphi^*(\bm{r}')}{1 + \varphi(\bm{r}')\varphi^*(\bm{r}')}
 \right]
\nonumber \\
&&
+ \frac{i\eta c_{\rm dd}F}{2\hbar} \varphi^2 \int d^2 r' 
n_{\rm tot}(\bm{r}') \sum_{\bm{k}} e^{i\bm{k}\cdot (\bm{r}-\bm{r}')}
\left[ 
- h_1(k)
\frac{\varphi^*(\bm{r}')}{1 + \varphi(\bm{r}')\varphi^*(\bm{r}')}
+ h_2^*(k)
\frac{\varphi(\bm{r}')}{1 + \varphi(\bm{r}')\varphi^*(\bm{r}')}
 \right]
\nonumber \\
&&
+ \frac{i\eta c_{\rm dd}F}{\hbar} \varphi \int d^2 r' 
n_{\rm tot}(\bm{r}') 
\sum_{\bm{k}} e^{i\bm{k}\cdot (\bm{r}-\bm{r}')} h_1(k)
\frac{1 - \varphi(\bm{r}')\varphi^*(\bm{r}')}
{1 + \varphi(\bm{r}')\varphi^*(\bm{r}')},
\label{phiA.0}
\end{eqnarray}
where the subscript $\perp$ was omitted for simplicity and 
\begin{eqnarray}
 h_1(k) &=& \frac{8\pi}{3} - 4\pi G(kd),
\\
 h_2(k) &=&  4\pi G(kd) (\hat{k}_x + i\hat{k}_y)^2.
\end{eqnarray}

For case (ii), we take $\hat{e}_{\rm n}=\hat{y}$, $\hat{e}^B=\hat{z}$, and
$p=0$, and use Eqs. (\ref{eq:b_bar}), (\ref{f.1.2D}), 
(\ref{eq:Q_rot_2D_k}), and (\ref{Qk.B}). 
Then, the equation of motion of spins is described as
\begin{eqnarray}
 \frac{\partial\varphi(\bm{r},t)}{\partial t} 
&=& 
-\bm{v}_{\rm mass}\cdot\nabla\varphi
+ \frac{i\hbar}{2Mn_{\rm tot}}\nabla n_{\rm tot}\cdot\nabla\varphi
+ \frac{i\hbar}{2M}\nabla^2\varphi
- \frac{i\hbar}{M} \frac{\varphi^*(\nabla\varphi)^2}{1+\varphi\varphi^*}
\nonumber \\
&&
- \frac{i\eta c_{\rm dd}F}{\hbar} \int d^2 r' 
n_{\rm tot}(\bm{r}') 
\sum_{\bm{k}} e^{i\bm{k}\cdot(\bm{r} - \bm{r}')}
\tilde{\mathcal{Q}}_{\bm{k}}
\frac{\varphi(\bm{r}')}{1 + \varphi(\bm{r}')\varphi^*(\bm{r}')}
\nonumber \\
&&
+ \frac{i\eta c_{\rm dd}F}{\hbar} \varphi^2 \int d^2 r' 
n_{\rm tot}(\bm{r}') 
\sum_{\bm{k}} e^{i\bm{k}\cdot(\bm{r} - \bm{r}')}
\tilde{\mathcal{Q}}_{\bm{k}}
\frac{\varphi^*(\bm{r}')}{1 + \varphi(\bm{r}')\varphi^*(\bm{r}')}
\nonumber \\
&&
- \frac{2i\eta c_{\rm dd}F}{\hbar} \varphi \int d^2 r'
n_{\rm tot}(\bm{r}')
\sum_{\bm{k}} e^{i\bm{k}\cdot(\bm{r} - \bm{r}')}
\tilde{\mathcal{Q}}_{\bm{k}}
\frac{1 - \varphi(\bm{r}')\varphi^*(\bm{r}')}
{1 + \varphi(\bm{r}')\varphi^*(\bm{r}')}
\nonumber \\
&&
+ \frac{iq(2F-1)}{\hbar} \frac{1 - \varphi\varphi^*}{1 + \varphi\varphi^*}
\varphi,
\label{phiB.0}
\end{eqnarray}  
where 
$\bm{r}_\perp$ $\to$ $\bm{r}=(x,z)$ and $\bm{k}_\perp$ $\to$ 
$\bm{k}=(k_x,k_z)$. 

\section{\label{sec.instability} Dynamical instability}

The hydrodynamic equations derived above give a rather straightforward
approach to the analysis of the spin dynamics in a spinor BEC.
In this section, we analyze the dynamical instability for cases (i) and
(ii). Here we consider a uniform quasi-2D system and assume 
$\nabla n_{\rm tot}=0$. 

\subsection{\label{sec.inst_i}Case (i): Instability under zero external field}

Here we analyze the dynamical instability under zero external field for two
initial stationary structures: uniform spin structures polarized 
normal to the $x$-$y$ plane ($\varphi_0 = 0$) and
in the $x$-$y$ plane ($\varphi_0 = 1$).  

First, we consider the case in which the spins are polarized normal to the
$x$-$y$ plane, i.e., in the $z$ direction, $\varphi_0 = 0$. 
Substituting $\varphi = 0 + \delta\varphi$ and 
$\bm{v}_{\rm mass} = \bm{v}_0 + \delta\bm{v}$ into
Eq.~(\ref{phiA.0}), we obtain linearized equations of 
$\delta\varphi$ and $\delta\varphi^*$. 
Performing Fourier expansions 
$\delta\varphi = \sum_{\bm{k}}\delta\tilde{\varphi}_{\bm{k}} 
e^{i\bm{k}\cdot\bm{r}}$ and 
$\delta\varphi^* = \sum_{\bm{k}}\delta\tilde{\varphi}^*_{-\bm{k}} 
e^{i\bm{k}\cdot\bm{r}}$,
we have
\begin{equation}
 \frac{d}{dt} \left(
\begin{array}{c}
 \delta\tilde{\varphi}_{\bm{k}} \\
 \delta\tilde{\varphi}^*_{-\bm{k}}
\end{array}
\right)
= \frac{i}{\hbar} \left(
\begin{array}{cc}
 - g_0 - g_1 & - g_2 \\
 g_2^* & - g_0 + g_1
\end{array}
\right) \left(
\begin{array}{c}
 \delta\tilde{\varphi}_{\bm{k}} \\
 \delta\tilde{\varphi}^*_{-\bm{k}}
\end{array}
\right),
\label{d-phi_k}
\end{equation}
where
\begin{eqnarray}
 g_0(\bm{k}) &=& \hbar \bm{v}_0\cdot\bm{k}, 
\label{g0.A0}
\\
 g_1(\bm{k}) &=& \frac{\hbar^2 k^2}{2M} 
- 2\pi\tilde{c}_{\rm dd} \left[ 2 - G(kd) \right],
\label{g1.A0}
\\
 g_2(\bm{k}) &=& 
2\pi\tilde{c}_{\rm dd} G(kd)(\hat{k}_x + i\hat{k}_y)^2. 
\label{g2.A0}
\end{eqnarray}
Here, $\tilde{c}_{\rm dd}=\eta c_{\rm dd} n_{\rm tot}F$ and 
$\bm{k}=(k_x,k_y)$. 
The eigenvalues of the $2 \times 2$ matrix in Eq.~(\ref{d-phi_k}) are
\begin{equation}
 \lambda_{\pm}(\bm{k}) = -\frac{i}{\hbar} g_0
\pm \frac{1}{\hbar} \sqrt{|g_2|^2 - g_1^2}.
\label{eq.lambda}
\end{equation}

\begin{figure}[tb]
\includegraphics[width=4cm]{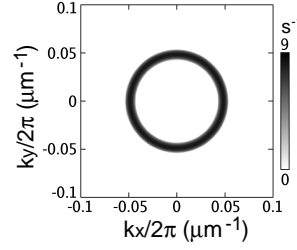}
\caption{\label{fig.i0} Re$\lambda_{+}(\bm{k})$
of the uniform spin structure polarized in the $z$ direction under zero
external field. 
The fluctuations in the black region are dynamically unstable and grow
 exponentially. 
Here, $n_{\rm tot}$ is given by $\sqrt{2\pi d^2}n_{\rm 3D}$ with 
$n_{3D}=2.3\times 10^{14}$ cm$^{-3}$ and $d=1.0$ $\mu$m.
The other parameters are given by the typical values for a spin-1
$^{87}$Rb atom: $M = 1.44 \times 10^{-25}$ Kg, $F=1$, $g_F = -1/2$, and 
$E_{\rm hf} = 6.835$ GHz $\times$ $h$.}
\end{figure}

The system becomes dynamically unstable when one of the eigenvalues has a
positive real part; that is when $\mathrm{Re} \lambda_{+}({\bm k})>0$ 
(or $|g_2|^2-g_1^2>0$). 
The wavevector dependence of $\mathrm{Re} \lambda_{+}$ is
shown in Fig.~\ref{fig.i0}. 
When the BEC is polarized perpendicular to the 2D plane, the MDDI is
repulsive and isotropic in the 2D plane. Thus, the BEC is unstable
against spin flip, and the unstable modes distribute isotropically in
the momentum space.  
The unstable region in the momentum space has a ring shape. The radius
and width of the ring are estimated as 
$k_0 = (2/\hbar)\sqrt{2\pi M\tilde{c}_{\rm dd}}$ 
and $\Delta k \simeq (\sqrt{\pi}/8)k_0^2d(4-\sqrt{\pi}k_0d)$,
respectively, for $kd\ll 1$. 
 
Next, we consider the uniform spin structure polarized in the $x$
direction, $\varphi_0 = 1$. 
We obtain the linearized equations in a similar way to the above.
Substituting $\varphi = 1 + \delta\varphi$ and 
$\bm{v}_{\rm mass} = \bm{v}_0 + \delta\bm{v}$ into
Eq.~(\ref{phiA.0}) and performing the Fourier expansion, 
we obtain the equation of the same form as Eq.~(\ref{d-phi_k}) with 
\begin{eqnarray}
 g_0(\bm{k}) &=& \hbar \bm{v}_0\cdot\bm{k}, 
\label{g0.A1}
\\
 g_1(\bm{k}) &=& \frac{\hbar^2 k^2}{2M} 
+ 2\pi\tilde{c}_{\rm dd} \left[ 1 - G(kd)(1 - \hat{k}^2_y) \right],
\label{g1.A1}
\\
 g_2(\bm{k}) &=& 
2\pi\tilde{c}_{\rm dd} \left[ 1 - G(kd)(1 + \hat{k}^2_y) \right].
\label{g2.A1}
\end{eqnarray}
In this case, it can be shown that $g_2^2-g_1^2$ is always negative.
Then, the eigenvalues which are given by the same form as
Eq. (\ref{eq.lambda}) are purely imaginary regardless of 
${\bm k}$. Hence, the spin-polarized state along the 2D plane is stable
under zero magnetic field. 

\subsection{\label{sec.inst_ii}Case (ii): Instability under a strong magnetic field}

Here we analyze the dynamical instability for the helical spin
structure,
$\varphi_0 = e^{i(\bm{\kappa}_\alpha\cdot\bm{r} - \omega_\alpha t)}$, 
characterized by the helix wavevector $\bm{\kappa}_\alpha$ in the
$x$-$z$ plane under a strong magnetic field in the $z$ direction.
Substituting $\varphi = \varphi_0$ and $\bm{v}_{\rm mass}=\bm{v}_0$
into Eq.~(\ref{phiB.0}) gives 
$\omega_\alpha = \bm{v}_0\cdot\bm{\kappa}_\alpha$.
Substituting $\varphi = \varphi_0(1 + \delta\varphi)$ and 
$\bm{v}_{\rm mass} = \bm{v}_0 + \delta\bm{v}$ into
Eqs.~(\ref{div.0}) and (\ref{eq.MHR.1}), and holding the terms up
to the first order of the fluctuations, we have 
$\nabla\cdot\bm{v}_0 = 0$, $\nabla\cdot\delta\bm{v} = 0$,
$\nabla\times\bm{v}_0 = 0$, and
$\nabla\times\delta\bm{v} = \frac{\hbar}{2M}
(\nabla\delta\varphi + \nabla\delta\varphi^*)\times\bm{\kappa}_\alpha$.
After Fourier expansions of $\delta\varphi$, $\delta\varphi^*$ and
 $\delta\bm{v} = \sum_{\bm{k}} \delta\tilde{\bm{v}}_{\bm{k}}
e^{i\bm{k}\cdot\bm{r}}$, we have
\begin{equation}
 \delta\tilde{\bm{v}}_{\bm{k}} = \frac{\hbar}{2M}
(\delta\tilde{\varphi}_{\bm{k}} + \delta\tilde{\varphi}^*_{-\bm{k}}) 
\left[
\bm{\kappa}_\alpha - \frac{(\bm{k}\cdot\bm{\kappa}_\alpha)\bm{k}}{k^2}
\right].
\label{d-vk}
\end{equation}
Substituting $\varphi = \varphi_0(1+\delta\varphi)$ 
into Eq.~(\ref{phiB.0}) and applying Eq.~(\ref{d-vk}), 
we obtain an equation of the same form as
Eq.~(\ref{d-phi_k}) with
\begin{eqnarray}
 g_0(\bm{k}) &=& \hbar \bm{v}_0\cdot\bm{k}, 
\label{g0.B}
\\
 g_1(\bm{k}) &=& \frac{\hbar^2 k^2}{2M} + \frac{\hbar^2}{2M} \left[
\frac{\kappa_\alpha^2}{2} - \frac{(\bm{k}\cdot\bm{\kappa}_\alpha)^2}{k^2}
\right] 
+ \frac{\tilde{c}_{\rm dd}}{4}
\left( \tilde{\mathcal{Q}}_{\bm{k} + \bm{\kappa}_\alpha} 
+ \tilde{\mathcal{Q}}_{\bm{k} - \bm{\kappa}_\alpha} \right)
- \tilde{c}_{\rm dd}
\left( \tilde{\mathcal{Q}}_{\bm{\kappa}_\alpha} 
+ \tilde{\mathcal{Q}}_{\bm{k}}  \right)
+ \frac{q(2F-1)}{2},
\label{g1.B}
\\
 g_2(\bm{k}) &=& \frac{\hbar^2}{2M} \left[
\frac{\kappa_\alpha^2}{2} - \frac{(\bm{k}\cdot\bm{\kappa}_\alpha)^2}{k^2}
\right] 
- \frac{\tilde{c}_{\rm dd}}{4}
\left( \tilde{\mathcal{Q}}_{\bm{k} + \bm{\kappa}_\alpha} 
+ \tilde{\mathcal{Q}}_{\bm{k} - \bm{\kappa}_\alpha} \right)
- \tilde{c}_{\rm dd} \tilde{\mathcal{Q}}_{\bm{k}}
+ \frac{q(2F-1)}{2}.
\label{g2.B}
\end{eqnarray}
Here, $\bm{k}=(k_x,k_z)$ and the Fourier
transform of the dipole kernel is now simply given by
$\tilde{\mathcal{Q}}_{\bm{k}} = (2\pi/3)[1 - 3 (k_z/k)^2G(kd)]$.
The eigenvalues of the $2\times 2$ matrix in Eq.~(\ref{d-phi_k})
are given by Eq.~(\ref{eq.lambda})
with Eqs.~(\ref{g0.B})--(\ref{g2.B}).
When $\bm{\kappa}_\alpha // \hat{z}$, $\bm{v}_0=0$, and neither the
MDDI nor the quadratic Zeeman effect  exists
($\tilde{c}_{\rm dd}=q=0$), the eigenvalues coincide with the
dispersion relation derived in Ref.~\cite{lama}. 

\begin{figure*}[tb]
\includegraphics[width=12cm]{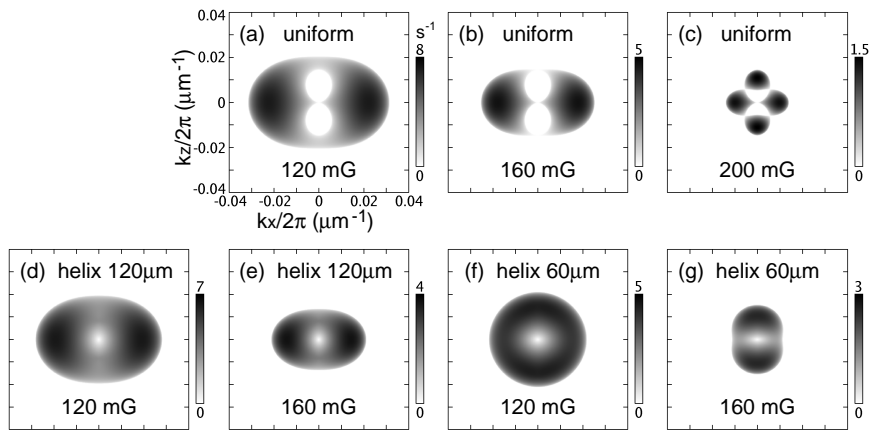}
\caption{\label{fig.hx} Re$\lambda_{+}(\bm{k})$ 
for (a)--(c) uniform spin structures and (d)--(g) spin helices
in the $z$ direction with a pitch $2\pi/\kappa_\alpha$
under a strong magnetic field. 
The helical pitch for (d) and (e) is $120$ $\mu$m, 
and that for (f) and (g) is $60$ $\mu$m.
The external field $B$ is [(a), (d), and (f)] 120 mG, 
[(b), (e), and (g)] 160 mG, and (c) 200 mG. 
The other parameters are the same as those in Fig.~\ref{fig.i0}.}
\end{figure*}

Figure~\ref{fig.hx} illustrates the wavenumber dependence of
Re$\lambda_{+}(\bm{k})$ for uniform and 
helical spin structures under various magnetic field strengths,
indicating the region of dynamically unstable modes.  
The dynamical instability discussed here agrees qualitatively
with that obtained by the Bogoliubov analysis~\cite{cherng,kawa09}. 
However, there is a quantitative discrepancy in the magnetic field
dependence of unstable modes caused by the fact that 
the local magnetization of the condensate is assumed to be fully
polarized in our method. 
However, when $q$ is not sufficiently small 
compared with the ferromagnetic
interaction, the amplitude of the magnetization decreases as $q$
increases. For the parameters used in the calculation for
Fig.~\ref{fig.hx}, our 
assumption is valid for $B \ll 480$ mG. 

\section{\label{sec.fluctuation} Magnetic fluctuation preference}

We also investigate the magnetic fluctuation preference for two cases,
that of dynamical instability under zero field for the uniform spin
structure polarized normal to the $x$-$y$ plane ($\varphi_0=0$), which
is discussed in Sec.~\ref{sec.inst_i}, and 
that of dynamical instability 
under a strong magnetic field for the helical spin structure 
($\varphi_0=e^{i(\bm{\kappa}_\alpha\cdot\bm{r}-\omega_\alpha t)}$),
which is discussed in Sec.~\ref{sec.inst_ii}.

For the case of zero external field for $\varphi_0=0$, two kinds of
magnetic fluctuations are considered: the $x$-direction fluctuation
$\delta\hat{f}_x$ and the $y$-direction fluctuation $\delta\hat{f}_y$.
They are described by the first order of $\delta\varphi$: 
$\delta\hat{f}_x=2\mathrm{Re}(\delta\varphi)$ and
$\delta\hat{f}_y=2\mathrm{Im}(\delta\varphi)$. Namely, the $x$-
and $y$-direction fluctuations are characterized by the real and
imaginary parts of $\delta\varphi$, respectively.

Let us reconsider the Fourier expansion of $\delta\varphi$,
\begin{eqnarray}
 \delta\varphi &=& \frac12 \sum_{\bm{k}\ne 0} 
(\delta\tilde{\varphi}_{\bm{k}} e^{i\bm{k}\cdot\bm{r}}
+ \delta\tilde{\varphi}_{-\bm{k}} e^{-i\bm{k}\cdot\bm{r}}) 
+ \delta\tilde{\varphi}_0
\nonumber \\
&=&  \frac12 \sum_{\bm{k}\ne 0} 
[A_{\rm R} \sin (\bm{k}\cdot\bm{r} + \alpha_{\rm R})
+ i A_{\rm I} \sin (\bm{k}\cdot\bm{r} + \alpha_{\rm I})] 
+ \delta\tilde{\varphi}_0,
\end{eqnarray}
where
\begin{eqnarray}
 A_{\rm R} &=& \sqrt{
[\mathrm{Re}(\delta\tilde{\varphi}_{\bm{k}} 
+ \delta\tilde{\varphi}_{-\bm{k}})]^2 + 
[\mathrm{Im}(\delta\tilde{\varphi}_{\bm{k}} 
- \delta\tilde{\varphi}_{-\bm{k}})]^2 },
\label{AR}
\\
 A_{\rm I} &=& \sqrt{
[\mathrm{Im}(\delta\tilde{\varphi}_{\bm{k}} 
+ \delta\tilde{\varphi}_{-\bm{k}})]^2 + 
[\mathrm{Re}(\delta\tilde{\varphi}_{\bm{k}} 
- \delta\tilde{\varphi}_{-\bm{k}})]^2 },
\label{AI}
\end{eqnarray}
and $\alpha_{\rm R}= \tan^{-1}[
\mathrm{Re}(\delta\tilde{\varphi}_{\bm{k}} 
+ \delta\tilde{\varphi}_{-\bm{k}})/
\mathrm{Im}(\delta\tilde{\varphi}_{-\bm{k}} 
- \delta\tilde{\varphi}_{\bm{k}})]$ and
 $\alpha_{\rm I}= \tan^{-1}[
\mathrm{Im}(\delta\tilde{\varphi}_{\bm{k}} 
+ \delta\tilde{\varphi}_{-\bm{k}})/
\mathrm{Re}(\delta\tilde{\varphi}_{\bm{k}} 
- \delta\tilde{\varphi}_{-\bm{k}})]$.
As $\lambda_{+}(\bm{k}=0)=0$, $\delta\tilde{\varphi}_0=0$. 
We introduce the quantity $\theta$, which characterizes the magnetic
fluctuation preference:
\begin{equation}
 \theta = \tan^{-1}(A_{\rm R}/A_{\rm I}).
\label{eq.theta.0}
\end{equation}

Here, we calculate $\theta(\bm{k})$ for the unstable modes shown in
Fig.~\ref{fig.i0}. The eigenvector which corresponds to the eigenvalue
$\lambda_{+}$ of the $2\times 2$ matrix in 
Eq.~\eqref{d-phi_k} is given by
\begin{equation}
\left(
\begin{array}{c}
 \delta\tilde{\varphi}_{\bm{k}} \\ \delta\tilde{\varphi}^*_{-\bm{k}}
\end{array}
\right)
=
\left(
\begin{array}{c}
 (g_1 + i\sqrt{|g_2|^2-g_1^2})/\sqrt{2|g_2|^2} \\
-g_2^*/\sqrt{2|g_2|^2}
\end{array}
\right),
\label{eq.eigvec}
\end{equation}
where $g_1$ and $g_2$ are defined by Eqs.~\eqref{g1.A0} and
\eqref{g2.A0}.
Then, from Eqs.~(\ref{AR})--(\ref{eq.eigvec}), we obtain
\begin{equation}
 \theta(\bm{k}) = \tan^{-1} \left[ \frac
{|g_2|^2 - g_1\mathrm{Re}(g_2) + \mathrm{Im}(g_2)\sqrt{|g_2|^2-g_1^2}}
{|g_2|^2 + g_1\mathrm{Re}(g_2) - \mathrm{Im}(g_2)\sqrt{|g_2|^2-g_1^2}}
\right]^{1/2}.
\label{eq.theta_k.A}
\end{equation}
We can consider the $x$-direction fluctuation to be
dominant for $\pi/4 < \theta < \pi/2$ and the $y$-direction fluctuation 
to be dominant for $0 < \theta < \pi/4$. 

\begin{figure}[tb]
\includegraphics[width=12cm]{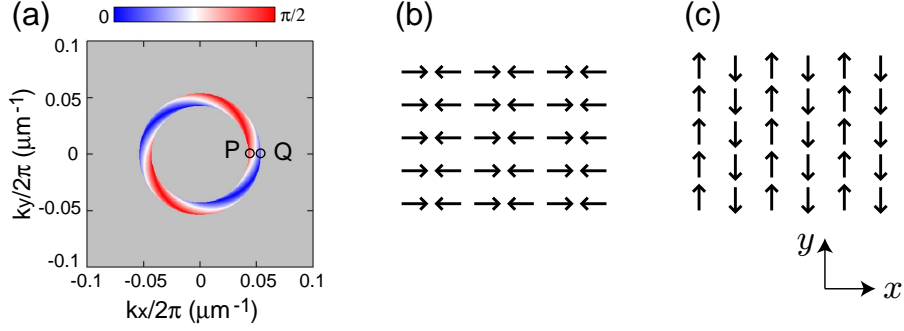}
\caption{\label{fig.nu0} (Color) (a) Magnetic fluctuation preference
 $\theta(\bm{k})$ for the unstable mode shown in Fig.~\ref{fig.i0}. 
The $x$-direction fluctuation is dominant in the red regions 
($\pi/4 < \theta < \pi/2$) and the $y$-direction fluctuation is dominant
 in the  blue regions ($0 < \theta < \pi/4$).
(b)(c) Schematic pictures of the magnetic patterns induced by the
 dynamical instability at 
 points (b) P and (c) Q designated in (a).
The arrows show the local magnetizations projected onto the $x$-$y$ plane.
The MDDI energy for configuration (c) is lower than that for (b).
}
\end{figure}

The wavevector dependence of $\theta(\bm{k})$ for the unstable mode
shown in Fig.~\ref{fig.i0} is illustrated in Fig.~\ref{fig.nu0}(a). 
In the red (blue) regions, 
the $x$($y$)-direction fluctuation is more dominant  than
the $y$($x$)-direction fluctuation. 
The schematic pictures of the magnetic patterns induced by the dynamical
instability at points P and Q are illustrated in 
Figs.~\ref{fig.nu0}(b) and (c), respectively.
At point P in Fig.~\ref{fig.nu0}(a), $x$-direction
fluctuation occurs and the wavevector of the magnetic pattern is
directed in the
$x$ direction as shown in Fig.~\ref{fig.nu0}(b).  
The MDDI energy for this configuration is higher than that of
Fig.~\ref{fig.nu0}(c), where $y$-direction fluctuation
is induced. 
In other words, the MDDI energy is reduced by the magnetic fluctuation,
and the reduction is larger in pattern (c) than in pattern (b).
The reduction in MDDI energy is converted into kinetic energy
($\sim k^2$), which is higher at point Q than at point P in
Fig.~\ref{fig.nu0}(a).   This explains why the magnetic
fluctuations change from $x$-direction to $y$-direction along the $k_x$
axis. 

Now, we investigate the magnetic fluctuations for the situation
discussed in Sec.~\ref{sec.inst_ii}. The fluctuations are considered to be
longitudinal or transverse.  
The longitudinal and transverse fluctuations are represented by 
$\delta\hat{f}_z$ and $\delta(\hat{f}_x + i\hat{f}_y)$, respectively.
Substituting $\varphi = \varphi_0 (1 + \delta\varphi)$ and
$|\varphi_0|^2 = 1$ into Eq.~(\ref{f_xyz}), we obtain the expressions for
the two types of fluctuations described by the first order of
$\delta\varphi$: 
$\delta\hat{f}_z = -\mathrm{Re} (\delta\varphi)$ and 
$\delta(\hat{f}_x + i\hat{f}_y) = \varphi_0 \mathrm{Im} (\delta\varphi)$.
Namely, the longitudinal and transverse fluctuations are characterized
by the real and imaginary parts of $\delta\varphi$, respectively.

We can also apply the above method to discuss the magnetic
fluctuation preference, which is characterized by
$\theta=\tan^{-1}(A_{\rm R}/A_{\rm I})$, in the present case. 
Since $g_2=g_2^*$, we
can simplify Eq.~\eqref{eq.theta_k.A} as
\begin{equation}
\theta(\bm{k}) = \tan^{-1} \sqrt{\frac{g_2- g_1}{g_2+ g_1}},
\label{eq.theta_k.B}
\end{equation}
where $g_1$ and $g_2$ are defined by Eqs.~\eqref{g1.B} and \eqref{g2.B},
respectively. The magnetic
fluctuation is longitudinal if $\pi/4<\theta<\pi/2$ and transverse if 
$0<\theta<\pi/4$.

\begin{figure}[tb]
\includegraphics[width=7cm]{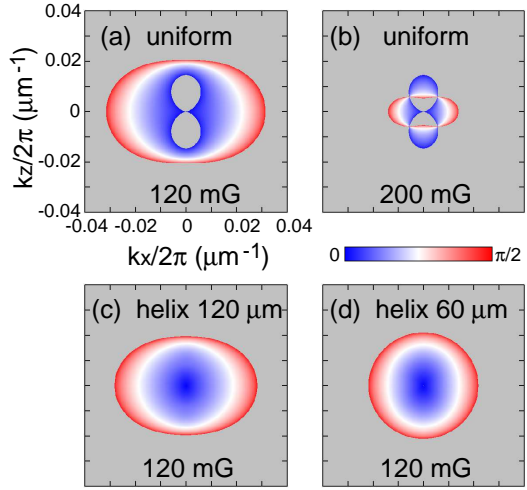}
\caption{\label{fig.nu} (Color) Magnetic fluctuation preference
 $\theta(\bm{k})$ for uniform and helical spin structures 
 in the $z$ direction with a pitch $2\pi/\kappa_\alpha$ $\mu$m under a
 strong magnetic field. 
The initial structure is uniform for (a) and (b),
a helix with a pitch (c) $120$ $\mu$m, and (d) $60$ $\mu$m.
The external field $B$ is [(a), (c), and (d)] 120 mG and (b) 200 mG. 
The fluctuations are longitudinal in the red regions ($\pi/4<\theta<\pi/2$) and
 transverse in the blue regions ($0<\theta<\pi/4$). The other parameters are
 the same as  those in Fig.~\ref{fig.hx}.
}
\end{figure}

The wavevector dependence of $\theta(\bm{k})$ for the unstable modes shown
in Figs.~\ref{fig.hx} (a), (c), (d), and (f) are demonstrated in
Fig.~\ref{fig.nu}. 
When the dynamical instability has a round shape [Figs.~\ref{fig.nu}(a),
(c), and (d)], the fluctuations are transverse for small $k$ and 
longitudinal for large $k$. 
Figure~\ref{fig.nu}(b) looks
more complex than the schematics for the other cases: 
fluctuations are transverse
for $\bm{k}//\hat{z}$ 
and longitudinal for $\bm{k}//\hat{x}$.
The magnetic fluctuation preference is consistent with that obtained by
the Bogoliubov analysis~\cite{kawa09, cherng}, and discrepancies appear 
in strong fields for the same reason as that for the dynamical
instability. 
 
\section{\label{sec.conc} Conclusions}

Employing our hydrodynamic description derived in Sec.~\ref{sec.hydro}, 
we have demonstrated some simple examples of the analysis of 
dynamical instability and magnetic fluctuation preference in
Sec.~\ref{sec.instability} and Sec.~\ref{sec.fluctuation}, respectively.
Once one finds a stationary solution of the hydrodynamic equations, 
it is a straightforward task to obtain the analytical form of the
dynamical instability: the only necessary step is the
diagonalization of a $2 \times 2$ matrix. The eigenvalues and
eigenvectors of the matrix lead to the dynamical instability and the
magnetic fluctuation preference, respectively.
Although we have discussed just a few types of spin structures and
external fields for simplicity,  
our method can be applied to other conformations.  

In conclusion, we have introduced the hydrodynamic equations for a
ferromagnetic spinor dipolar BEC with an arbitrary spin by means of
stereographic projection.   
This simple description provides a straightforward approach by which
to investigate spin dynamics, i.e.,
dynamical instability and magnetization
fluctuation preference, which are expressed in analytical forms.
The description should also be useful for the study of 
the exact solutions of hydrodynamic equations of a spinor BEC.

\begin{acknowledgments}
 The authors thank M. Ueda for his useful comments. 
This work is supported
by MEXT JSPS KAKENHI (No. 22103005, 22340114, 22740265), the Photon Frontier
Network Program of MEXT, Japan, Hayashi Memorial Foundation for Female Natural
Scientists, and JSPS and FRST under the Japan-New
 Zealand Research Cooperative Program.  
\end{acknowledgments}

\appendix

\section{\label{sec.A} Contributions from the short-range interaction, 
MDDI, and linear and quadratic Zeeman effects}

The contribution from the short-range interaction to the equation of motion of $
f_z$ is calculated as
\begin{eqnarray}
\left[ \frac{\partial f_z}{\partial t} \right]_{\rm s}
&=&
\frac{1}{i\hbar} (F_z)_{mn} \left[ 
\left( i\hbar \frac{\partial \Psi_m^*}{\partial t}\right) \Psi_n
+ \Psi_m^* \left( i\hbar \frac{\partial \Psi_n}{\partial t}\right)
\right]_{\rm s}
\nonumber \\
&=&
\sum_{S=0,{\rm even}}^{2F} \frac{4\pi\hbar}{iM} a_S 
\sum_{M_S=-S}^S \sum_{lm'l'} (F_z)_{mn}
\left[ -\langle ml | SM_S\rangle \langle S M_S|m'l'\rangle 
\Psi_l\Psi_{m'}^*\Psi_{l'}^*\Psi_n
\right. \nonumber\\
&&\hspace{50mm} \left.
+ \Psi_m^*\langle nl | SM_S\rangle \langle S M_S|m'l'\rangle 
\Psi_l^*\Psi_{m'}\Psi_{l'} \right]\nonumber\\
&=&\sum_{S=0,{\rm even}}^{2F} \frac{4\pi\hbar}{iM} a_S
\sum_{M_S=-S}^S\sum_{lm'l'}  
m\langle ml | SM_S\rangle \langle S M_S|m'l'\rangle 
\left[ -\Psi_m\Psi_l\Psi_{m'}^*\Psi_{l'}^*
+\Psi_m^*\Psi_l^*\Psi_{m'}\Psi_{l'} \right]\nonumber\\
&=&\sum_{S=0,{\rm even}}^{2F} \frac{4\pi\hbar}{iM}a_S 
\sum_{M_S=-S}^S\sum_{mm'}
(m-m')\langle m,M_S-m | SM_S\rangle \langle S M_S|m',M_S-m'\rangle 
\Psi_m^*\Psi_{M_S-m}^*\Psi_{m'}\Psi_{M_S-m'}\nonumber\\
&=&\sum_{S=0,{\rm even}}^{2F}\frac{2\pi\hbar}{iM}a_S \sum_{M_S=-S}^S
\bigg[ \sum_{mm'}
(m-m')\langle m,M_S-m | SM_S\rangle \langle S M_S|m',M_S-m'\rangle 
\Psi_m^*\Psi_{M_S-m}^*\Psi_{m'}\Psi_{M_S-m'}\nonumber\\
&&\hspace{40mm}+\sum_{ll'}
(-l+l')\langle M_S-l,l | SM_S\rangle \langle S M_S|M_S-l',l'\rangle 
\Psi_{M_S-l}^*\Psi_{l}^*\Psi_{M_S-l'}\Psi_{l'} \bigg]\nonumber\\
&=&0,
\end{eqnarray}
where we have used $(F_z)_{mn}=m\delta_{mn}$.
Here, $[\cdots]_{\rm s}$ denotes that only those terms that come from 
the short-range interaction are extracted.
In the following, this notation is applied to the contributions from the
MDDI ($[\cdots]_{\rm dd}$),
linear ($[\cdots]_p$), and quadratic ($[\cdots]_q$) Zeeman effects.
Since the short-range interaction~\eqref{eq.Vs} 
is invariant under spin rotation, 
$[\partial f_x/\partial t]_{\rm s}$ and $[\partial f_y/\partial t]_{\rm s}$ 
also vanish, which are shown in a similar way by choosing the spin 
quantization axis along the $x$ and $y$ directions, respectively.

The contribution from the MDDI to the equation of motion
of spin is calculated as
\begin{eqnarray}
\left[ \frac{\partial f_\mu}{\partial t} \right]_{\rm dd}
&=& 
\frac{1}{i\hbar} (F_\mu)_{mn} \left[ 
\left( i\hbar \frac{\partial \Psi_m^*}{\partial t}\right) \Psi_n
+ \Psi_m^* \left( i\hbar \frac{\partial \Psi_n}{\partial t}\right)
\right]_{\rm dd}
\nonumber \\
&=&
\frac{c_{\rm dd}}{i\hbar} (F_\mu)_{mn} \left[
- b_\nu^* (F_\nu^*)_{ml} \Psi_l^* \Psi_n
+ \Psi_m^* b_\nu (F_\nu)_{nl} \Psi_l
\right]
\nonumber \\
&=&
\frac{c_{\rm dd}}{i\hbar} b_\nu
\Psi_m^* (F_\mu F_\nu - F_\nu F_\mu)_{mn} \Psi_n
\nonumber \\
&=&
\frac{c_{\rm dd}}{i\hbar} b_\nu
i \epsilon_{\mu\nu\lambda} f_\lambda
\nonumber \\
&=&
\frac{c_{\rm dd}}{\hbar} (\bm{b}\times\bm{f})_\mu.
\end{eqnarray}
We have used the relations $F_\mu^{\dagger} = F_\mu$ and 
$[F_\mu, F_\nu] = i \epsilon_{\mu\nu\lambda}F_\lambda$.

Suppose the magnetic field is applied parallel to the $z$ axis.
Then, the contribution from the linear Zeeman effect is calculated as
\begin{eqnarray}
  \left[ \frac{\partial f_\mu}{\partial t} \right]_p
&=& 
\frac{1}{i\hbar} (F_\mu)_{mn} \left[ 
\left( i\hbar \frac{\partial \Psi_m^*}{\partial t}\right) \Psi_n
+ \Psi_m^* \left( i\hbar \frac{\partial \Psi_n}{\partial t}\right)
\right]_p
\nonumber \\
&=&
\frac{p}{i\hbar} (F_\mu)_{mn} \left[
- (F_z^*)_{ml} \Psi_l^* \Psi_n + \Psi_m^* (F_z)_{nl} \Psi_l
\right]
\nonumber \\
&=&
\frac{p}{i\hbar} i \epsilon_{\mu z \nu} f_\nu
\nonumber \\
&=&
\frac{p}{\hbar} (\hat{z}\times\bm{f})_\mu.
\end{eqnarray}

The contribution from the quadratic Zeeman effect is calculated in the
same way as the that shown above.
\begin{eqnarray}
  \left[ \frac{\partial f_\mu}{\partial t} \right]_q
&=& 
\frac{1}{i\hbar} (F_\mu)_{mn} \left[ 
\left( i\hbar \frac{\partial \Psi_m^*}{\partial t}\right) \Psi_n
+ \Psi_m^* \left( i\hbar \frac{\partial \Psi_n}{\partial t}\right)
\right]_q
\nonumber \\
&=&
\frac{q}{i\hbar} (F_\mu)_{mn} \left[
- (F_z^*)^2_{ml} \Psi_l^* \Psi_n + \Psi_m^* (F_z)^2_{nl} \Psi_l
\right]
\nonumber \\
&=&
\frac{q}{i\hbar} \Psi_m^* 
(F_z[F_\mu, F_z] + [F_\mu, F_z]F_z)_{mn} \Psi_n
\nonumber \\
&=&
\frac{q}{i\hbar} i\epsilon_{\mu z \nu}
\Psi_m^* (F_\nu F_z + F_z F_\nu)_{mn} \Psi_n
\nonumber \\
&=&
\frac{2q}{\hbar} n_{\rm tot} \epsilon_{\mu z \nu} \hat{\mathcal{N}}_{z\nu},
\end{eqnarray}
where $\hat{\mathcal{N}}_{\mu\nu}$ is a nematic tensor defined by 
\begin{equation}
 \hat{\mathcal{N}}_{\mu\nu} = \frac12 
\zeta_m^* (F_\mu F_\nu + F_\nu F_\mu)_{mn} \zeta_n.
\end{equation}


\begin{thebibliography}{99} 
\bibitem{Ho1996}
	T.-L. Ho and V. B. Shenoy,
	Phys. Rev. Lett. {\bf 77}, 2595 (1996).
\bibitem{Nakahara2000}
	M. Nakahara, T. Isoshima, K. Machida, S. Ogawa, and T. Ohmi,
	Physica B: Condensed Matter {\bf 284--288}, 17 (2000);
	T. Isoshima, M. Nakahara, T. Ohmi, and K. Machida,
	Phys. Rev. A {\bf 61}, 063610 (2000).
\bibitem{Leanhardt2003}
	A. E. Leanhardt, Y. Shin, D. Kielpinski, D. E. Pritchard, and W. Ketterle,
	Phys. Rev. Lett. {\bf 90}, 140403 (2003).

\bibitem{Makela}
	H. M\"akel\"a, Y. Zhang, and K.-A. Suominen,
	J. Phys. A: Math. Gen. {\bf 36}, 8555 (2003);
	H. M\"akel\"a, 
	J. Phys. A: Math. Gen. {\bf 39}, 7423 (2006).
\bibitem{Zhou2001}
	F. Zhou, 
	Phys. Rev. Lett. {\bf 87}, 080401 (2001).
\bibitem{Semenoff2007}
	G. W. Semenoff and F. Zhou,
	Phys. Rev. Lett. {\bf 98}, 100401 (2007).
\bibitem{Kobayashi2009}
	M. Kobayashi, Y. Kawaguchi, M. Nitta, and M. Ueda,
	Phys. Rev. Lett. {\bf 103}, 115301 (2009).

\bibitem{Sadler2006}
	L. E. Sadler, J. M. Higbie, S. R. Leslie, M. Vengalattore, and D. M. Stamper-Kurn,
	Nature {\bf 443}, 312 (2006).
\bibitem{berkeley08} M. Vengalattore, S. R. Leslie, J. Guzman, and
	D. M. Stamper-Kurn, Phys. Rev. Lett. {\bf 100}, 170403 (2008).
\bibitem{Vengalattore2010} 
	M. Vengalattore, J. Guzman, S. R. Leslie, F. Serwane, and D. M. Stamper-Kurn, 
	Phys. Rev. A {\bf 81}, 053612 (2010).

\bibitem{MDDI_review}
	For review of dipolar BECs, see
	T. Lahaye, C. Menotti, L. Santos, M. Lewenstein and T. Pfau,
	Rep. Prog. Phys. {\bf 72}, 126401 (2009).

\bibitem{Santos2006}
	L. Santos and T. Pfau,
	Phys. Rev. Lett. {\bf 96}, 190404 (2006).
\bibitem{Kawaguchi2006}
	Y. Kawaguchi, H. Saito and M. Ueda,
	Phys. Rev. Lett. {\bf 96}, 080405 (2006);
	Phys. Rev. Lett. {\bf 97}, 130404 (2006).
\bibitem{Yi2006}
	S. Yi and H. Pu,
	Phys. Rev. Lett. {\bf 97}, 020401 (2006).
\bibitem{kawa09} Y. Kawaguchi, H. Saito, K. Kudo, and M. Ueda,
	Phys. Rev. A {\bf 82}, 043627 (2010). 
\bibitem{Takahashi2007}
	M. Takahashi, Sankalpa Ghosh, T. Mizushima, and K. Machida,
	Phys. Rev. Lett. {\bf 98}, 260403 (2007);
	J. A. M. Huhtam\"aki, M. Takahashi, T. P. Simula, T. Mizushima, and K. Machida,
	Phys. Rev. A {\bf 81}, 063623 (2010).
\bibitem{lama} 
	A. Lamacraft, Phys. Rev. A {\bf 77}, 063622 (2008).
\bibitem{Barnett2009}
	R. Barnett, D. Podolsky, and G. Refael,
	Phys. Rev. B {\bf 80}, 024420 (2009).
\bibitem{Lamacraft2010}
	A. Lamacraft,
	Phys. Rev. B {\bf 81}, 184526 (2010).
\bibitem{Ho1998}
	T.-L. Ho, Phys. Rev. Lett. {\bf 81}, 742 (1998).
\bibitem{Mermin} N.D. Mermin and T.-L. Ho, Phys. Rev. Lett. {\bf 36},
	594 (1976).
\bibitem{LLG} S. Zhang and Z. Li, Phys. Rev. Lett. {\bf 93}, 
127204 (2004);
A. Thiaville, Y. Nakatani, J. Miltat, and Y. Suzuki, 
Europhys. Lett. {\bf 69}, 990 (2005).
 
\bibitem{kawa07} Y. Kawaguchi, H. Saito, and M. Ueda,
	Phys. Rev. Lett. {\bf 98}, 110406 (2007).
\bibitem{laksh} M. Lakshmanan and K. Nakamura, Phys. Rev. Lett. {\bf
	53}, 2497 (1984).
\bibitem{cherng} R.W. Cherng and E. Demler, Phys. Rev. Lett. {\bf 103},
	185301 (2009).
\end{thebibliography}
\end{document}